\documentclass[twocolumn]{aastex6}

\newcommand\obj{2015 RR$_{245}$}

\shorttitle{\obj\ in the 9:2 resonance}
\shortauthors{Bannister et al.}

\begin{document}

\title{OSSOS: IV. Discovery of a dwarf planet candidate in the 9:2 resonance with Neptune}


\author{Michele T. Bannister\altaffilmark{1}, 
Mike Alexandersen\altaffilmark{5}, 
Susan D. Benecchi\altaffilmark{7},
Ying-Tung Chen\altaffilmark{5}, 
Audrey Delsanti\altaffilmark{8}, 
Wesley C. Fraser\altaffilmark{9}, 
Brett J. Gladman\altaffilmark{4}, 
Mikael Granvik\altaffilmark{10}, 
Will M. Grundy\altaffilmark{12}, 
Aur\'elie Guilbert-Lepoutre\altaffilmark{3}, 
Stephen D. J. Gwyn\altaffilmark{2}, 
Wing-Huen Ip\altaffilmark{14, 15},
Marian Jakubik\altaffilmark{16},
R. Lynne Jones\altaffilmark{17}, 
Nathan Kaib\altaffilmark{18},
J. J. Kavelaars\altaffilmark{1,2}, 
Pedro Lacerda\altaffilmark{9}, 
Samantha Lawler\altaffilmark{2},
Matthew J. Lehner\altaffilmark{5, 20, 21}, 
Hsing Wen Lin\altaffilmark{14}, 
Patryk Sofia Lykawka\altaffilmark{23}, 
Michael Marsset\altaffilmark{8, 24},
Ruth Murray-Clay\altaffilmark{25}, 
Keith S. Noll\altaffilmark{26}, 
Alex Parker\altaffilmark{27},
Jean-Marc Petit\altaffilmark{3}, 
Rosemary E. Pike\altaffilmark{1,5}, 
Philippe Rousselot\altaffilmark{3}, 
Megan E. Schwamb\altaffilmark{31},
Cory Shankman\altaffilmark{1}, 
Peter Veres\altaffilmark{30},
Pierre Vernazza\altaffilmark{8}, 
Kathryn Volk\altaffilmark{6}, 
Shiang-Yu Wang\altaffilmark{5},
Robert Weryk\altaffilmark{29}
}
\email{michele.t.bannister@gmail.com}
\altaffiltext{1}{Department of Physics and Astronomy, University of Victoria, Elliott Building, 3800 Finnerty Rd, Victoria, BC V8P 5C2, Canada}
\altaffiltext{2}{NRC-Herzberg Astronomy and Astrophysics, National Research Council of Canada, 5071 West Saanich Rd, Victoria, BC V9E 2E7, Canada}
\altaffiltext{3}{Institut UTINAM UMR6213, CNRS, Univ. Bourgogne Franche-Comt\'e, OSU Theta F25000 Besan\c{c}on, France}
\altaffiltext{4}{Department of Physics and Astronomy, University of British Columbia, Vancouver, BC, Canada}
\altaffiltext{5}{Institute for Astronomy \& Astrophysics, Academia Sinica; 11F AS/NTU, National Taiwan University, 1 Roosevelt Rd., Sec. 4, Taipei 10617, Taiwan}
\altaffiltext{6}{Department of Planetary Sciences/Lunar \& Planetary Laboratory, University of Arizona, 1629 E University Blvd, Tucson, AZ 85721, USA}
\altaffiltext{7}{Planetary Science Institute, 1700 East Fort Lowell, Suite 106, Tucson, AZ 85719, USA}
\altaffiltext{8}{Aix Marseille Universit\'e, CNRS, LAM (Laboratoire d'Astrophysique de Marseille) UMR 7326, 13388, Marseille, France}
\altaffiltext{9}{Astrophysics Research Centre, Queen's University Belfast, Belfast BT7 1NN, United Kingdom}
\altaffiltext{10}{Department of Physics, P.O. Box 64, 00014 University of Helsinki, Finland}
\altaffiltext{12}{Lowell Observatory, Flagstaff, Arizona, USA}
\altaffiltext{14}{Institute of Astronomy, National Central University, Taiwan}
\altaffiltext{15}{Space Science Institute, Macau University of Science and Technology, Macau}
\altaffiltext{16}{Astronomical Institute, Slovak Academy of Science, 05960 Tatranska Lomnica, Slovakia}
\altaffiltext{17}{University of Washington, Washington, USA}
\altaffiltext{18}{HL Dodge Department of Physics \& Astronomy, University of Oklahoma, Norman, OK 73019, USA}
\altaffiltext{20}{Department of Physics and Astronomy, University of Pennsylvania, 209 S. 33rd St., Philadelphia, PA 19104, USA}
\altaffiltext{21}{Harvard-Smithsonian Center for Astrophysics, 60 Garden St., Cambridge, MA 02138, USA}
\altaffiltext{23}{Astronomy Group, School of Interdisciplinary Social and Human Sciences, Kindai University, Japan}
\altaffiltext{24}{European Southern Observatory (ESO), Alonso de C\'ordova 3107, 1900 Casilla Vitacura, Santiago, Chile}
\altaffiltext{25}{Department of Physics, University of California, Santa Barbara, CA 93106, USA}
\altaffiltext{26}{NASA Goddard Space Flight Center, Code 693, Greenbelt, MD 20771, USA}
\altaffiltext{27}{Southwest Research Institute, Boulder, Colorado, USA}
\altaffiltext{29}{Institute for Astronomy, University of Hawaii, 2680 Woodlawn Drive, Honolulu HI 96822, USA}
\altaffiltext{30}{Jet Propulsion Laboratory, California Institute of Technology, Pasadena, CA 91109, USA}
\altaffiltext{31}{Gemini Observatory, Northern Operations Center, 670 North A'ohoku Place, Hilo, HI 96720, USA}

\begin{abstract}

We report the discovery and orbit of a new dwarf planet candidate, 2015 RR$_{245}$, 
by the Outer Solar System Origins Survey (OSSOS).
\obj's orbit is eccentric ($e$=0.586), with a semi-major axis near 82~au, 
yielding a perihelion distance of 34~au.
\obj\ has $g-r = 0.59 \pm 0.11$ and absolute magnitude $H_{r} = 3.6 \pm 0.1$; 
for an assumed albedo of $p_V = 12$\% the object has a diameter of $\sim670$~km.
Based on astrometric measurements from OSSOS and Pan-STARRS1, we find that
2015 RR$_{245}$ is securely trapped on ten-Myr timescales in the 9:2 mean-motion resonance
with Neptune. It is the first TNO identified in this resonance.
On hundred-Myr timescales, particles in \obj-like orbits depart 
and sometimes return to the resonance, 
indicating that \obj\ likely forms part of the long-lived metastable 
population of distant TNOs
that drift between resonance sticking and actively scattering via 
gravitational encounters with Neptune. 
The discovery of a 9:2 TNO stresses the role of 
resonances in the long-term evolution of objects in the 
scattering disk, and reinforces the view that distant resonances 
are heavily populated in the current Solar System. 
This object further motivates detailed modelling of the transient sticking population. 
\end{abstract}


\section{Introduction} \label{sec:intro}

The Outer Solar System Origins Survey (OSSOS) was designed to provide 
a set of $500+$ very precise trans-Neptunian object (TNO) orbits by the end of its 
2013--2017 observations with the Canada-France-Hawaii Telescope 
(CFHT) \citep{Bannister:2015}. 
As OSSOS covers 155 square degrees of sky on and near the Solar
System mid-plane, the Kuiper belt's steep luminosity function 
\citep{2001AJ....122.1051G, Petit:2011p3938, Fraser:2014vt}
was used to predict that the brightest target expected to be found over the course of the 
survey would have apparent magnitude $m_r\sim 21.5$.  
At $m_r=21.8$, \obj\ is the brightest target discovered by OSSOS.
At a current heliocentric distance of 65 au, this bright OSSOS
detection is also far beyond the median distance of TNO detections in sky surveys.
Its substantial distance requires \obj\ to be sizeable.

\section{Discovery and Size}

\obj\ was discovered with apparent mean magnitude $m_{r} = 21.76 \pm 0.01$
in three images taken with CFHT MegaCam in an $r$-band filter over a two-hour span on 9 September 2015 (Table~\ref{tab:photometry}).
The TNO was found within the 21 deg$^{2}$ survey region 
centred at RA $0^{h}30^{m}$, Declination $+5.0^{\circ}$, 
the seventh of the eight OSSOS survey areas.
The discovery analysis was as described in \citet{Bannister:2015}.
\obj\ is a characterized discovery within OSSOS: 
its discovery magnitude is within the range where
the survey has a calibrated detection efficiency and full tracking of all discoveries to a high-precision orbit was possible.
A full de-biasing of the region's discoveries will be accomplished using the OSSOS survey simulator \citep{Bannister:2015} in the future. 
Because OSSOS reduces an entire discovery semester of data after the 
semester's observations are complete, the software first yielded the object 
in late January 2016.
OSSOS imaging is designed to provide tracking throughout the discovery semester; 
these data quickly yielded further astrometry in September through December 2015.

In February 2016, just before \obj\ moved close to solar conjunction, a
sequential pair of images in $g$ and $r$ (Table~\ref{tab:photometry}) yielded a preliminary broadband colour of 
$g-r = 0.59 \pm 0.11$.
The relatively neutral colour (cf. solar $g-r = 0.44\pm0.02$\footnote{\url{http://www.sdss.org/dr12/algorithms/ugrizvegasun/}})
 is more common to a dynamically excited 
``hot'' Kuiper Belt object rather than that found in the cold classical 
belt \citep{Doressoundiram:2005haa, Peixinho:2015bw}.

We re-observed \obj\ in early June 2016 
as part of planned OSSOS recovery observations for the fields observed in 2015. 
\obj\ was found to have $m_r \sim 21.8$ (Table~\ref{tab:photometry}), 
consistent with the brightness found in the September 2015 discovery observations.
All measurements can be retrieved from the IAU Minor Planet Center (MPC)\footnote{\url{http://www.minorplanetcenter.net/db_search/show_object?utf8=\%E2\%9C\%93&object_id=2015+RR245}};
photometry referred to here is summarized in Table~\ref{tab:photometry}.

Based on the heliocentric distance (see \S~\ref{sec:orbit}), 
the absolute magnitude $H_r$ is $3.6 \pm 0.1$ ($H_V = 3.8 \pm 0.1$\footnote{$H_V = H_r - 0.03 + 0.45(g-r)$ \citep{2002AJ....123.2121S}}).
The albedos for $2 < H_{V} < 4$ TNOs that have been determined by thermal 
measurements range between $p_V = 7$\% (2002 MS$_{4}$; $H_{V}=4.0$) and 21\% (Quaoar; $H_{V}=2.7$) \citep{Brucker:2009jp}.
For an assumed albedo $p_V$ at each end of this range, less or more 
reflective respectively, \obj's diameter is 870--500 km.
For an albedo of 9\% like that of $H_{V} = 2.0$ TNO 2007 OR$_{10}$ 
\citep{2016arXiv160303090P},
which is on a comparable orbit ($a$=67 au, $q$=33 au) at a current distance of $88$ au \citep{Schwamb:2010p932},
\obj's diameter would be 770 km. 
The neutral colour of \obj\ leans it towards being part of the neutral colour class, 
which has lower albedos of $\sim6$\% \citep{Lacerda:2014wr}.
However, due to the wide range of albedos seen for objects in this 
$H_{V}$ range \citep{Lellouch:2013cu}, we adopt a modal albedo of 
$p_V = 12$\% (D$\sim 670$ km) in the rest of this discussion.

\begin{deluxetable*}{llcllD}
\tablecaption{Selected observations with CFHT MegaCam of \obj\ \label{tab:photometry}}
\tablehead{
\colhead{Time (UT)} & \colhead{Filter} & \colhead{Exposure} \vspace{-0.07cm} & \colhead{Magnitude} & \colhead{IQ ('')}  &  \colhead{Target} \vspace{-0.07cm} \\
\colhead{} & \colhead{} & \colhead{Time (s)} & \colhead{} & \colhead{} & \colhead{Elongation ($^\circ$)} 
}
\startdata
2015 09 09.37654   & R.MP9602	& 300	&  $21.77 \pm 0.02$	&	0.42	&	159.3 \\
2015 09 09.42092   & R.MP9602	& 300	&  $21.77 \pm 0.02$	&	0.42	&	159.3 \\
2015 09 09.46188   & R.MP9602	& 300	&  $21.73 \pm 0.02$ 	&	0.42	&	159.4 \\
2016 02 04.22299   & R.MP9602	& 200	&  $22.04 \pm 0.06$ 	&	1.21	&	51.4 \\
2016 02 04.22679  & G.MP9402	& 200	&  $22.63 \pm 0.07$ 	&	1.56	&	51.4	\\
2016 06 07.61302  & R.MP9602	& 300	&  $21.94 \pm 0.03$ 	&	0.55	& 	69.5 \\
2016 06 08.57085  & R.MP9602	& 300	&  $21.88 \pm 0.02$ 	&	0.60	&	70.4 \\
\enddata
\tablecomments{The photometry is calibrated to the SDSS per the methodology in \citet{Bannister:2015}. 
The filter bandpasses are similar to those of Sloan.
}
\end{deluxetable*}

The plausible diameter range for \obj\ is interesting 
because it spans the range of sizes where significant changes occur in TNO surface composition, 
particularly with the presence of deep water ice absorption \citep{Brown:2012cv}.
At this size scale, objects with the expected ice/rock compositional 
mix predominant in the outer Solar System are expected to adjust to an
approximately spherical hydrostatic equilibrium shape 
\citep{2008Icar..195..851T, Lineweaver:2010te}.
The majority of the possible diameter range places \obj\ in the size range where 
self-gravity produces a spherical shape.
By this criterion it could be considered one of the roughly 
20--30 ``dwarf planet candidates" now known from previous wide-field 
shallow surveys covering most of the sky 
\citep{Larsen:2007fd, Trujillo:2003p584, Schwamb:2010p932, 2011AJ....142...98S, Rabinowitz:2012eq, Bannister:2013tn, Brown:2015dt}, most of which have $a<50$~au.

\section{Orbit}
\label{sec:orbit}

With all available OSSOS astrometry from September 2015 to June 2016,
the distance to \obj\ could be accurately evaluated as $64.479 \pm 0.008$~au.
However, the barycentric orbital semi-major axis was less precisely determinable: 
$a=82.1\pm2.5$~au (1-sigma covariance-based error estimate).
With a swath of semi-major axis uncertainty more than five au wide, it 
was impossible to confidently determine resonance occupation: 
typical resonance widths are 0.5--1~au in the outer Solar System.
Without a very precise orbit, multiple dynamical behaviours are possible,
and this object could not usefully advance the discussion of the subtleties 
of resonance sticking below.
At this point the object was released to the MPC so
that the worldwide community could participate in the second year of
recovery and provide physical characterization.
On July 13, 2016, the Pan-STARRS1 survey \citep{Kaiser:2010gr}
released six oppositions of astrometry of \obj\ to the MPC, 
spanning Sept 2010 to July 2015.
A subset of these observations were found independently by the analysis of \citet{2016arXiv160704895W};
they were augmented with additional Pan-STARRS1 detections found based on the \citet{2016arXiv160704895W} orbit.

\begin{figure*}[ht]
\figurenum{1}
\plotone{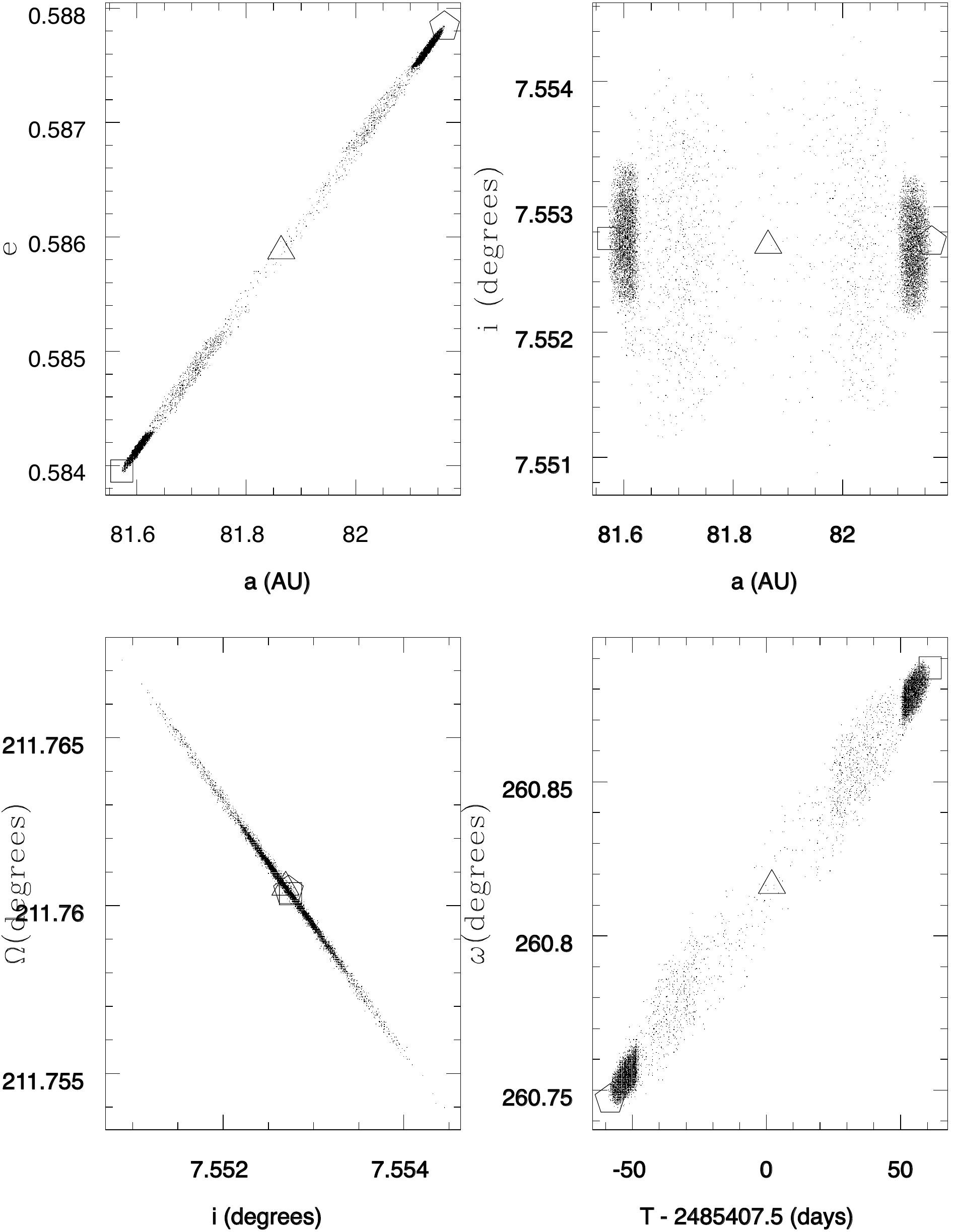}
\caption{Uncertainties on the orbit of 2015 RR$_{245}$ using the accounting 
for possible systematics fully described in \citet{Gladman2008}.
The best-fit orbit (Table~\ref{tab:orbit}) is marked by the large open triangle in each
panel, while the square and pentagon give orbital elements corresponding 
to the minimal and maximal semi-major axes (respectively) of the Monte Carlo
search.  
Small dots show orbits consistent with the available astrometry
found during the search for the two extremal orbits; their density 
is not proportional to likelihood and are not used elsewhere in 
this manuscript.
All elements are barycentric in the J2000 reference frame and thus judged
relative to the ecliptic.
(upper left) Semi-major axis $a$ versus eccentricity $e$.  Note the strong
correlation that is equivalent to a set of orbits having nearly the same
$q$=33.9~au perihelion distance as the best fit orbit.
(upper right) Ecliptic inclination $i$ as a function of $a$.
(lower left) Longitude of ascending node $\Omega$ as a function of $i$.
(lower right) Argument of perihelion $\omega$ versus the Julian
day of osculating perihelion judged at the epoch JD= 2457274.9.
\label{fig:orbit}}
\end{figure*}

\begin{figure}[ht]
\figurenum{2}
\plotone{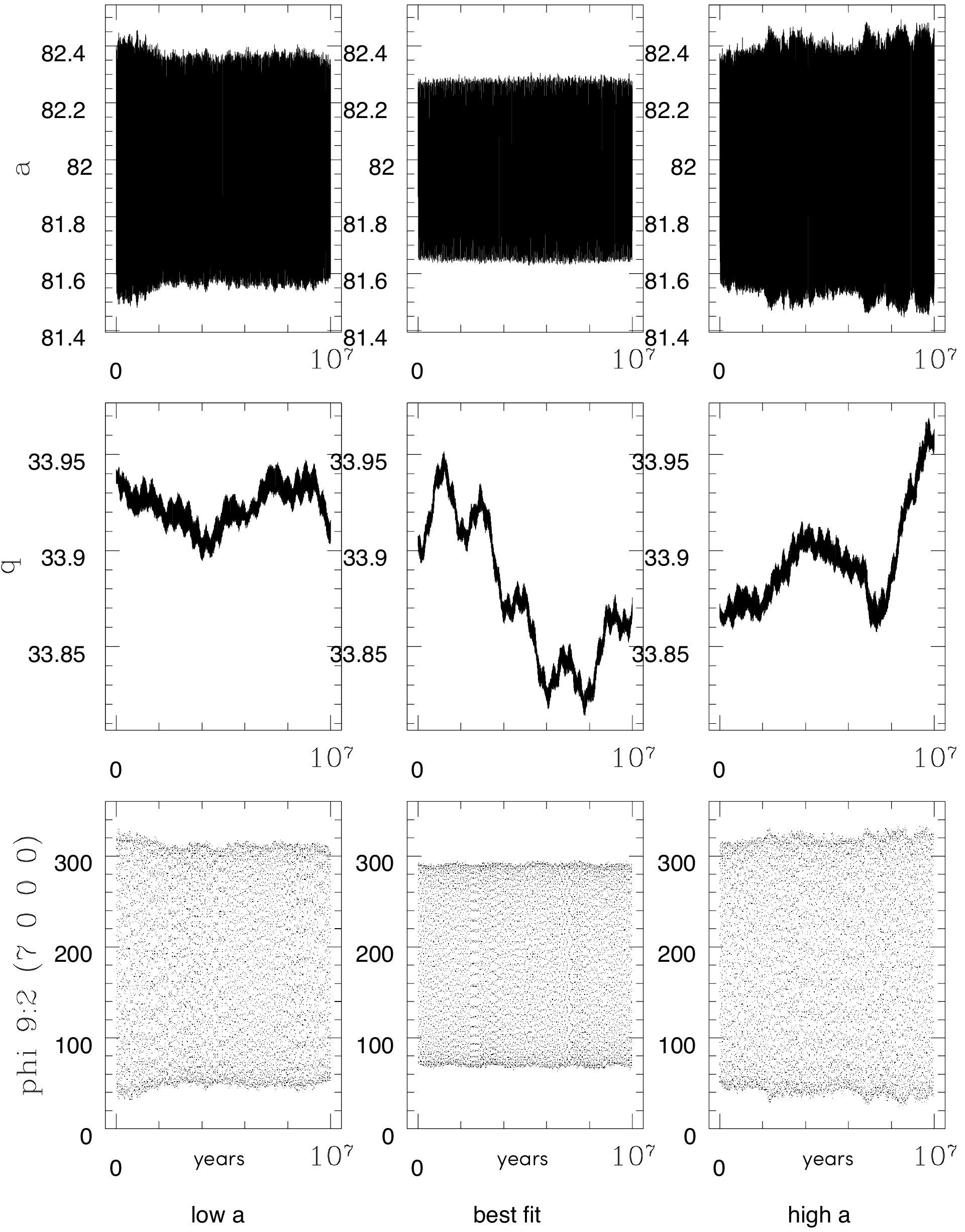}
\caption{The orbital evolution of the three orbits marked in Fig.~\ref{fig:orbit}.
The left, middle, and right columns show the $a$, $q$, and resonant
argument $\Phi$ evolution for the minimal, best-fit, and maximal semi-major axis orbits.
The $\simeq 0.5$~au oscillation of $a$ is forced by the resonance and is
coupled to the rapid ($\simeq$10,000 year) libration of the resonant argument.
The dynamical protection provided by the resonance results in only very
weak interactions allowing only very slow evolution of the perihelion distance $q$.
The libration amplitude in the resonance is $\simeq110^\circ$ for
the best fit orbit and is 20--30 degrees larger for the extremal orbits.
\label{fig:evolution}}
\end{figure}

The orbital solution to the seven-opposition set of astrometric measurements \citep[calculated per][]{Bernstein:2000p444}
provides a secure classification when analyzed with 
the dynamical orbital classification algorithm of \citet{Gladman2008}:
all plausible orbits yield the same dynamical behaviour.
The best-fit J2000 barycentric orbital elements are given in
Table~\ref{tab:orbit}.
Fig.~\ref{fig:orbit} shows the large span of all plausible 
orbital elements obtained by the Metropolis algorithm described by 
\citet{Gladman2008}.
The span of plausible orbital behaviours gives a secure dynamical 
classification in the 9:2 mean-motion resonance with Neptune 
(Fig.~\ref{fig:evolution}), which is 
centred\footnote{Note that this evaluation should be made in
barycentric orbital element space. Neptune's barycentric semi-major axis 
$a_N=30.07$ au is perturbed by up to 0.02 au on 500-Myr time scales. 
For instance, while the barycentric mean centre of the resonance is shifting 
on the 10 Myr interval in Fig.~\ref{fig:evolution} (upper row), 
it is not apparent due to being $\sim1/200^{th}$ the barycentric oscillation of the particles.}
at a barycentric semi-major axis of 81.96 au.
\obj\ is the first trans-Neptunian object securely identified in this distant resonance.

\begin{deluxetable*}{llllll}
\tablecaption{Barycentric elements for \obj\ in ICRS at osculating epoch 2457274.9 \label{tab:orbit}}
\tablehead{
\colhead{$a$ (au)} & \colhead{$e$} & \colhead{$i$ ($^{\circ}$)} & \colhead{$\Omega$ ($^{\circ}$)} & \colhead{$\omega$ ($^{\circ}$)} & \colhead{JD peri}
}
\startdata
81.86 $\pm$ 0.05  & 0.5859 $\pm$ 0.0003  &  7.553 $\pm$ 0.001  &  211.761 $\pm$ 0.002  &  260.817 $\pm$ 0.012  & 2485409 $\pm$ 10 \\
\enddata
\tablecomments{Barycentric distance: 64.479 $\pm$ 0.001 au with true anomaly $f = 253.5^\circ$.  \\ 
$\Omega$ is the longitude of ascending node, 
$\omega$ the argument of perihelion, 
and {\it JD peri} is the Julian day of osculating barycentric perihelion. 
The uncertainties are the $1\sigma$ estimates based on the covariance matrix 
at the best-fit orbit, derived by the method of \citet{Bernstein:2000p444}. 
}
\end{deluxetable*}

This resonance occupancy reinforces the finding that there are many TNOs 
in high-order, distant $a > 50$ resonances 
\citep{Chiang:2003hb, Lykawka:2007ff, Gladman2008, Gladman:2012ed, Alexandersen:2014uv, Pike:2015gn, Sheppard:2016tx, Kaib:2016tz}.
Objects in large-$a$ resonances are inefficiently discovered
due to the $r^{-4}$ dependence for reflected flux, the overall steep 
TNO luminosity function, and because the large eccentricities of 
such orbits places most of the population at large distances at any given time,
and thus below the flux limit of wide-field surveys. 
When carefully debiased for detectability, the large-$a$ resonances together yield 
a total resonant population that is comparable to the main classical Kuiper belt 
\citep{Gladman:2012ed, Pike:2015gn, Volk:2015}.

\begin{figure}[ht]
\figurenum{3}
\plotone{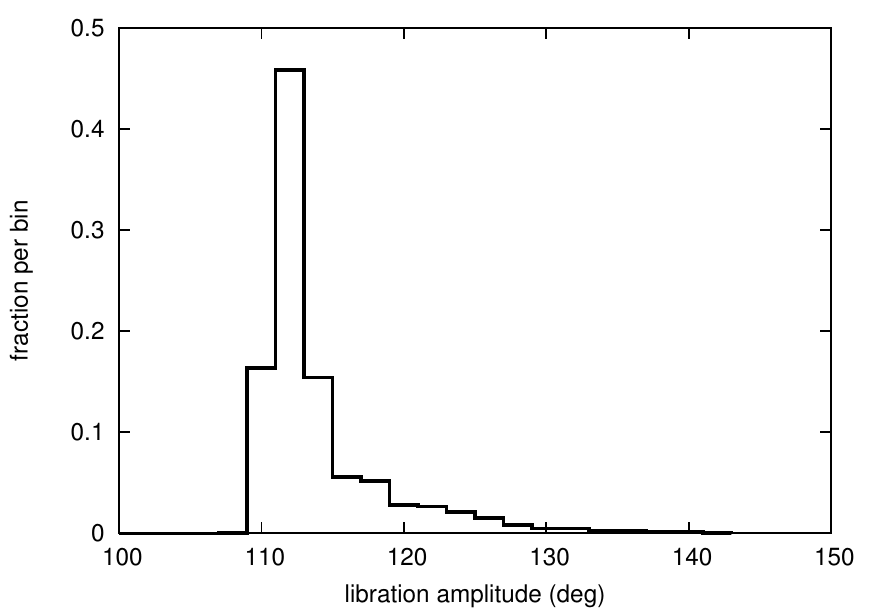}
\caption{Histogram of the $L_{92}$ libration amplitude distribution of the 250
clones produced by the covariance matrix analysis.
See text for discussion.
\label{fig:libamp}}
\end{figure}

The resonant protection provided by the 9:2 resonance is very similar
to that provided by the 3:2 and 5:2 mean-motion resonances 
\citep{1965AJ.....70...10C, Gladman:2012ed}.
Specifically, libration of the resonant argument
$\Phi_{92} = 9 \lambda - 2 \lambda_N - 7 \varpi $
around $180^\circ$ (Fig.~2) means that when a resonant TNO is at 
perihelion, Neptune is never near the same mean orbital longitude 
$\lambda_N$, preventing close encounters (in this expression,
$\lambda$ is the mean longitude of the particle and $\varpi = \Omega + \omega$
is the longitude of perihelion).
The libration amplitude $L_{92}$ results in the perihelion longitude
offset between the TNO and Neptune varying from $90^\circ - L_{92}/2$
to $90^\circ + L_{92}/2$.
Using methods described in \citet{Volk:2015}, the $L_{92}$ distribution 
(Fig.~\ref{fig:libamp}) was determined with 10 Myr simulations of 250
particles (`clones') distributed probabilistically in the $3 \sigma$
error ellipse card on the covariance matrix based on the orbit best-fit. 
All these orbits remained resonant for the 10 Myr duration; at the
current epoch, \obj\ is thus firmly lodged in the resonance.

There are many resonant TNOs with precisely known orbits that are stable for the lifetime of the 
Solar System, Pluto being an obvious example.
There also exist TNOs with similar high-precision orbits whose orbital evolution shows 
many resonant librations for all clones initially, but then have nearly 
all clones leave the resonance on time scales $\ll$4~Gyr.
Several Neptune 1:1 resonators (Trojans) are known to exhibit this
phenomenon \citep{Horner:2012ix, Alexandersen:2014uv}, 
as do three well-studied TNOs in the 5:1 \citep{Pike:2015gn}.
It is thus important to keep in mind the distinction between
(1) actively `scattering' TNOs, 
(2) `temporarily' metastable TNOs, 
(3) stable resonators and 
(4) permanently detached TNOs with no chance of recoupling on solar-system time scales; 
orbital resonances are often involved when TNOs transit between those states.

\begin{figure}
\figurenum{4}
\plotone{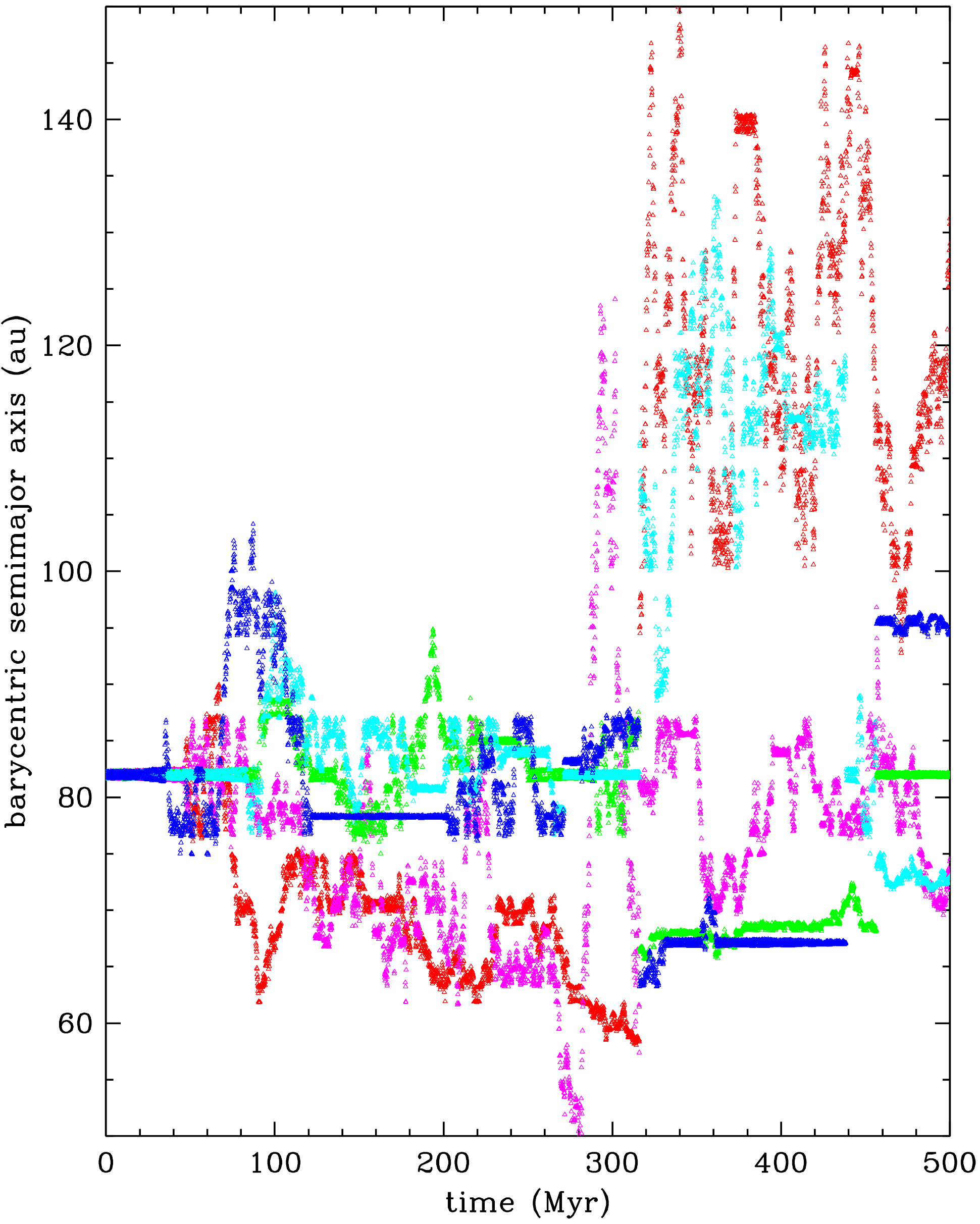}
\caption{Semi-major axis evolution for 5 sample orbits 
(each a different colour)
drawn from a covariance-based set of 250 orbits consistent
with the astrometry of \obj.
For the first 40 Myr, all 250 clones oscillate by the
$\simeq1$~au full width of the resonance.
They then begin leaving the resonance into
the actively scattering population where their semi-major axes 
are changing by several au/Myr in a diffusive manner driven by Neptune scattering.
The median time before first departure from the resonance is $\simeq100$~Myr; 
here 5 orbits with less typical early departures are shown, 
sometimes displaying the frequently seen behaviour among 
the clone ensemble of subsequent sticking to other resonances or
return to the 9:2.
}
\label{fig:longterm}
\end{figure}

To explore the longer-term evolution of \obj,
we extended the integration of these same 250 clones to 500 Myr.
While every orbit initially spends at least 10 Myr steadily librating
in the resonance, the clones begin to diffuse out on time scales of
50~Myr and begin actively scattering due to Neptune
(Fig.~\ref{fig:longterm}).
Their subsequent evolution then becomes diffusive, with the scattering
particles sometimes temporarily sticking to other resonances (during which
the evolution shows a stable semi-major axis at some other value). 
Particles commonly return to stick to the 9:2 resonance
itself, and then remain stuck for typical time scales of tens of Myr.
We find a median dynamical lifetime
before first departure from the resonance of order 100 Myr, with
roughly 15\% of the clones in the resonance 500 Myr in the future
(note that some may have left and returned; Fig.~\ref{fig:longterm}).
Only a tiny fraction of the clones will be in the 9:2 after 4 Gyr.
It is thus very likely that \obj\ has not continuously spent the 
last 4~Gyr in the resonance, but instead 
was trapped from the actively scattering population within the last 
$\sim$100 Myr.

\section{Discussion}
\label{sec:discussion}

Two main possibilities appear likely for how \obj\ came to be in its current orbit: first, that it was scattered off Neptune and is presently `sticking' to a resonance, with the scattering event either recent or early in Solar System history.
Secondly, \obj\ could have been captured into the resonance during 
Neptune's migration. We consider each in turn.

Metastable resonant TNOs that are emplaced by `transient sticking' are an established phenomenon.  
The transient sticking slows orbital evolution, providing a mechanism necessary to maintain the current scattering disk, 
which would otherwise decay on timescales much shorter than the age of the Solar System \citep{Duncan:1997hg}.
Several studies of transient sticking report temporary captures in the 9:2 resonance  \citep{Fernandez:2004kh, Lykawka:2007dj, Almeida:2009bib}.  
These studies found most periods spent in the resonance are short ($\sim 10$ Myr) 
and with large libration amplitudes $L_{92}>130^\circ$.
However, occasionally their modelled particles attained smaller libration amplitudes, 
which lengthened their occupation in the resonance.
The low-libration `sticker' objects provide an enhanced contribution to the steady-state transient population.  
Indeed, the simulations reported in \citet{Lykawka:2007dj} include stickers in the 9:2 resonance 
with $L_{92}$ as small as the $\simeq115^\circ$ observed for \obj. 
Of the particles in \citet{Lykawka:2007dj} surviving to the present time,
roughly half had experienced a trapping in the 9:2 at some time, 
and one case kept $L_{92}<120^{\circ}$ for 700~Myr (P. Lykawka, 2016 private communication).
\obj\ plausibly fits into this `metastable TNO' paradigm.

Given the $\sim$100 Myr median resonant lifetime of our orbital clones, we 
suggest that \obj\ is likely to be transiently alternating between Neptune 
mean-motion resonances and the actively scattering component of the trans-Neptunian region. 
This conclusion is bolstered by \obj's perihelion distance of $q=34$ au; 
roughly\footnote{Orbital integration is required; see discussions 
in \citet{Lykawka:2007ff} and \citet{Gladman2008}.}  
$q<37$~au results in the continuing orbital interactions with Neptune 
that are shared by almost all active scatterers.
In fact, non-resonant TNOs with $q$ only 4 au from Neptune's orbital semi-major axis of 30 au 
typically experience sufficiently numerous strong encounters with Neptune 
(when the longitude of Neptune matches that of the TNO while the TNO is at perihelion) 
for their orbits to rapidly evolve \citep{Morbidelli:2004dh}.  
Such rapidly evolving orbits, when observed at the current epoch, are classified in the scattering population \citep{Gladman2008}.
Residence in the 9:2 or other resonances can temporarily shield 
TNOs from scattering, but eventually their orbital evolution will lead 
such TNOs to leave the resonance and resume active scattering.
Note that there can be resonant objects (eg.~Pluto) which do not
participate at all in this process on 4~Gyr time scales.

Though we consider transient sticking the most plausible origin for \obj, 
other emplacement scenarios are possible.  
For example, Nice model-type histories \citep{2008Icar..196..258L}
could in principle emplace objects in the 9:2 resonance directly during an early Solar System upheaval event.  
If fortunate enough to remain stable for the subsequent $\sim$4 Gyr solar system age, these objects might still be present today.  
A numerical simulation of TNO sculpting under a Nice model-like Solar System history \citep{Pike2016nice} 
does produce a population of 9:2 resonant TNOs.
However, the objects reported in \citet{Pike2016nice},
drawn from the simulations of \citet{Brasser:2013dw},
 may also be transient captures. 
Further work is needed to determine whether these objects were caught early and retained, 
or whether they are in fact transiently sticking TNOs captured later in the 4 Gyr numerical evolution.  

Alternately, resonance capture during smooth migration of Neptune, even 
over a relatively large 10 au distance, would require an initial disk 
extending beyond 55 au to provide a source of TNOs for capture 
into resonance.  
While there are low-inclination TNOs beyond the 2:1 resonance that suggest 
that the cold classical TNO population did extend to at least 50 au \citep{Bannister:2015},
55 au would be at the larger end of the observed debris disk population \citep{Hilenbrand2008}.
Because they appeal to capture early in the Solar System's history, both the
Nice-type and the smooth migration scenarios would require that future observations 
of \obj\ push its orbit to a subset of phase space more stable than that 
currently explored by our orbital clones; 
this seems unlikely given the extent of our numerical exploration.

We next consider whether our detection of a large TNO in the resonant phase of the metastable population
is consistent with the population ratios between the two phases (scattering/resonant).
Our initial numerical experiments suggest  that---summed over all resonances---
the transiently stuck population may be comparable to the population of 
active scatterers.
This is similar to the behaviour seen for the known 5:1 resonant TNOs, 
which typically spend half their lifetimes in various resonances and half in a scattering state \citep{Pike:2015gn}.
If so, a single transiently-stuck dwarf planet candidate detection by OSSOS is consistent with 
our lack of detection of similarly-sized active scatterers.
Only a few of the well-populated distant resonances are known to contain 
$H<4$ TNOs \citep{2011AJ....142...98S}.
Additionally, the classification methods of \citet{Elliot:2005ju} and \citet{Gladman2008} 
both agree that with its current astrometric measurements, 
2007 OR$_{10}$ is securely in the 10:3 resonance.
Because dynamical timescales are longer at large semi-major axes, 
transiently stuck TNOs spend more time in more distant (low-order) resonances, 
making the 9:2 a reasonable resonance in which to find \obj.

When viewed in absolute magnitude $H$ space, detection of an $H_r=3.6$
TNO by OSSOS is naively a $\sim4$\% probability using the TNO sky density estimates of \citet[figure 9]{Fraser:2014vt}.
However, the $H_r$ frequency distribution reported in \citet{Fraser:2014vt}
utilizes an empirical formulation that adjusts for the 
increased albedos of many large TNOs
\citep{Brown:2008tp, Fraser:2008p103, Fraser:2014vt}.  
Use of that relation\footnote{The $H_r$ mag of the \obj\ is used to 
estimate a size given an estimated intrinsic albedo of $p_V = 12$\% 
and then an 'effective' $H_r$ mag is computed for that size using an 
effective albedo of 6\%.}
to compute an `effective' $H_r$ for \obj\ results in a value of $H_{r, eff} = 4.35$.
At this $H_{r, eff}$, our detection of one TNO in 155 square degrees of survey coverage 
is in good agreement with the measured $H_{r, eff}$ frequency 
distribution.

Concentrating on such `large' TNOs, \obj\ spends approximately two-thirds 
of its orbit brighter than the shallowest magnitude limit $m_r \sim 24.5$ of any OSSOS 
block; even at aphelion its sky motion of $\sim1"$/hour would be easily detectable by our survey.
With such a substantial visibility fraction, a trivial estimate of the number of comparable
TNOs within about 10 degrees of the ecliptic is $(360\times20 / 155) \simeq 50$ $H_r < 3.6$ 
TNOs over the sky, with only a small upward correction 
(of $<$50\%) for the fraction of the visibility.
Demanding that these TNOs be also in the 9:2 should be viewed as dangerous `post-facto'
reasoning (in that the argument would apply to any sub-population in which the single
TNO was found). 
Instead, the perspective should be that there are 50-100 $H_r < 3.6$ TNOs in the volume 
inside 100 au, which seems completely plausible.
Its dynamics suggest \obj\ is one of the objects that survived the population decay in the initially
scattered disk after experiencing scattering and temporary capture in multiple resonances.
If of order 100 $H\lesssim4$ TNOs exist and the ``retention efficiency" over
the entire outer Solar System is $\sim$1\% \citep{Duncan:1997hg, Nesvorny:2016dd}, 
then there would have been $\sim 10,000$ such objects present in 
the outer Solar System at the time that the giant planets began to 
clear the region. 
This is in line with primordial estimates
\citep{1991Icar...90..271S, 1997AJ....114..841S} of $\sim1000$ Plutos, 
when one takes into account that Pluto-scale TNOs are only a fraction of
the $H<4$ inventory.

Viewed another way, there may still be an issue due to the puzzling fact that
\obj\ is roughly 3 magnitudes brighter than the OSSOS detection limits.
That is, OSSOS detects many TNOs with $m_r<24.9$, and none
are in the 9:2 resonance.
We find that the $H_{v} = 4.7$ TNO 2003 UA$_{414}$,
recently re-found by Pan-STARRS,\footnote{Four new oppositions of observation in MPS 719376: \url{http://www.minorplanetcenter.net/iau/ECS/MPCArchive/2016/MPS_20160719.pdf}, which changed the semi-major axis of 2003 UA$_{414}$'s orbit by nearly a factor of two.} 
is securely in the 9:2, and stably resonant on a 100 Myr timescale, with 100 clones evolved as in \S~\ref{sec:orbit} all remaining resonant.
No other published surveys suggest any smaller TNOs being detected in the 9:2.
If one anchors a normal exponential magnitude distribution to \obj, even restricting
to its discovery distance of 65 AU, there should be $\sim100$ TNOs up to three magnitudes
fainter, yet only one has been found.
The problem is worsened when considering that near the $q=34$ au perihelion distance, 
TNOs as faint as $H_r\simeq9$ are visible to OSSOS, 
and detection of those TNOs is far more likely than finding \obj.
A plausible resolution of this apparent paradox is most likely that \obj\ 
has an albedo that is higher than that of smaller 
TNOs \citep{2008ssbn.book..161S}, as suggested above, 
and thus this TNO does not anchor a steep exponential distribution.
Considering known large TNOs on potentially `metastable' orbits,
for an albedo like that of the substantially larger Eris (at a current heliocentric distance of 96 au) 
of $p\simeq0.96$ \citep{Sicardy:2011gq},
the effective $H_r$ becomes nearly 6, 
and the non-detection of smaller TNOs even at perihelion is not statistically alarming.
We point out, however, that the 1500~km diameter 2007 OR$_{10}$'s visual 
albedo is only 9\% \citep{2016arXiv160303090P},
raising doubt on whether all large TNOs have high albedos \citep{Brown:2008tp}.
Future thermal measurements and spectral studies of \obj, which will steadily 
brighten as it approaches its 2090 perihelion,
will inform the open question of its albedo and surface composition.

\acknowledgments

This research was supported by funding from the National Research Council of Canada and the National Science and Engineering Research Council of Canada. 
The authors recognize and acknowledge the sacred nature of Maunakea, and appreciate the opportunity to observe from the mountain. 
This project could not have been a success without the dedicated staff of the Canada--France--Hawaii Telescope (CFHT) telescope. 
Based on observations obtained with MegaPrime/MegaCam, a joint project of CFHT and CEA/DAPNIA.
CFHT is operated by the National Research Council of Canada, the Institute National des Sciences de l'Universe of the Centre National de la Recherche Scientifique of France, and the University of Hawaii, 
with OSSOS receiving additional access due to contributions from the Institute of Astronomy and Astrophysics, Academia Sinica, Taiwan.
This work is based in part on data produced and hosted at the Canadian Astronomy Data Centre, with data processing and analysis performed using computing and storage capacity provided by the Canadian Advanced Network For Astronomy Research (CANFAR).
MES was supported by the Gemini Observatory, which is operated by the Association of Universities for Research in Astronomy, Inc., on behalf of the international Gemini partnership of Argentina, Brazil, Canada, Chile, and the United States of America. 
M.J. acknowledges support from the Slovak Grant Agency for Science (grant no. 2/0031/14).

\vspace{5mm}

\facilities{CFHT (MegaPrime), Pan-STARRS1}
\software{Python, astropy, matplotlib, scipy, numpy, supermongo, SWIFT}

\bibliography{rr245}
\bibliographystyle{apj}


\listofchanges

\end{document}